\documentclass[pra,twocolumn,floatfix]{revtex4}

\usepackage{graphicx}
\usepackage{amsmath}
\usepackage{xcolor}
\usepackage{ulem}
\usepackage[T1]{fontenc}
\usepackage{float}

\newcommand{\be}{\begin{equation}}
\newcommand{\ee}{\end{equation}}

\newcommand{\comment}[1]{}
\newcommand{\ket}[1]{\left|#1\right\rangle}
\newcommand{\bra}[1]{\left\langle#1\right|}
\newcommand{\ip}[2]{\left\langle#1\left|\right.#2\right\rangle}
\newcommand{\ev}[1]{\left\langle#1\right\rangle}

\newcommand{\g}{\text{g}}
\newcommand{\e}{\text{e}}
\definecolor{mypink1}{rgb}{0.858, 0.188, 0.478}
\definecolor{byzantium}{rgb}{0.44, 0.16, 0.39}
\definecolor{mygray}{gray}{0.6}

\begin{document}
\title{Dynamics of matter-wave quantum emitters in a structured vacuum}

\author{Michael Stewart}
\author{Joonhyuk Kwon}
\author{Alfonso Lanuza}
\author{Dominik Schneble}
\email[Please direct correspondence to ]{dominik.schneble@stonybrook.edu}
\affiliation{Department of Physics and Astronomy, Stony Brook University, Stony Brook, New York 11794-3800, USA}

\begin{abstract}
The characteristics of spontaneous emission can be strongly modified by the mode structure of the vacuum. In waveguide quantum-electrodynamics based on photonic crystals, this modification is exploited to engineer atom-photon interactions near a band edge, but the physics of coupling to an entire band has not yet been explored in experiments. Using ultracold atoms in an optical lattice, we study the decay dynamics of matter-wave quantum emitters coupled to a single band of an effective photonic crystal waveguide structure with tunable characteristics. Depending on the ratio between vacuum coupling and bandwidth, we observe a transition from irreversible decay to fully oscillatory dynamics linked to the interplay of matter-wave bound states near the band edges, whose spatial structure we characterize. Our results shed light on the emergence of coherence in an open quantum system in a controllable environment, and are of relevance for the understanding of vacuum-induced decay phenomena in photonic systems.
\end{abstract}
\maketitle

\section{Introduction}
Harnessing light-matter interactions is a central topic in the development of quantum technologies and  the emergent field of waveguide QED \cite{Cirac2017,Hood2016,Dibyendu2017}, where quantum emitters are coupled to strongly confined optical fields, opening up new avenues for the realization of photonic quantum matter in the optical and microwave domains \cite{Douglas2015,Chang2018,Carusotto2020}. 
A common approach, based on the use of photonic crystals (or band-gap materials) \cite{Yablonovich1987,John1987}, exploits their band structure and diverging density of states to enhance the coupling to guided photon modes. In the framework of cavity-QED \cite{Miller2005,Walther2006,Haroche2006}, such a band structure of guided modes can also be engineered in an array of coupled cavities \cite{Lombardo2014,Calajo2016}.

A fundamental question for these systems as a platform for  applications is the understanding of how quantum emitters interact with the modified vacuum. Spontaneous decay processes near a continuum edge \cite{Rzazewski1982,Kofman1994,Lambropoulos2000} are subject to the influence of an atom-photon bound state \cite{Bykov1975,John1990}, resulting in fractional decay of the excited state population. Such bound states and their effects have recently been explored in photonic \cite{Hood2016,Liu2017} and matter-wave platforms \cite{Krinner2018,GTudela2018C}. When the modified vacuum possesses a true band structure with multiple edges, another type of bound state is predicted to exist \cite{Lombardo2014,Calajo2016}. This secondary bound state can lead to qualitatively new physics, in that the dynamics now interpolates between Markovian decay and fully coherent oscillations as in the cavity-QED limit. When there are multiple bound states, a situation addressed in this work, their features are non-trivially influenced by the vacuum structure and thus may deviate from the simple exponential localization observed near a single edge \cite{Krinner2018}. As these bound states are proposed for engineering long-range atom-photon interactions \cite{Cirac2017,Hood2016,Liu2017,Sundareshan2019}, deviations from the predicted behavior should be relevant to studies of photonic-band-gap materials and waveguide-QED. In the following, we present an experimental study where we explore these questions in detail, based on a full and independent control over the coupling-to-bandwidth ratio and the excitation energy, as well as an effectively infinite Purcell factor with negligible coupling to modes other than those of interest.

These studies are made possible by a recently developed experimental platform \cite{Stewart2017,Krinner2018} that implements an array of matter-wave quantum emitters  in an optical lattice \cite{Vega2008} in which ultracold atoms take the role of single photons in the analogous photonic context. While the free-space emission of matter waves \cite{Krinner2018} is equivalent to the emission of photons near a zero-momentum band edge, we now create a structured vacuum for matter-wave emission in full analogy to that provided by a photonic crystal using an optical lattice.

\begin{figure*}[ht!]
\centering
    \includegraphics[width=1.9\columnwidth]{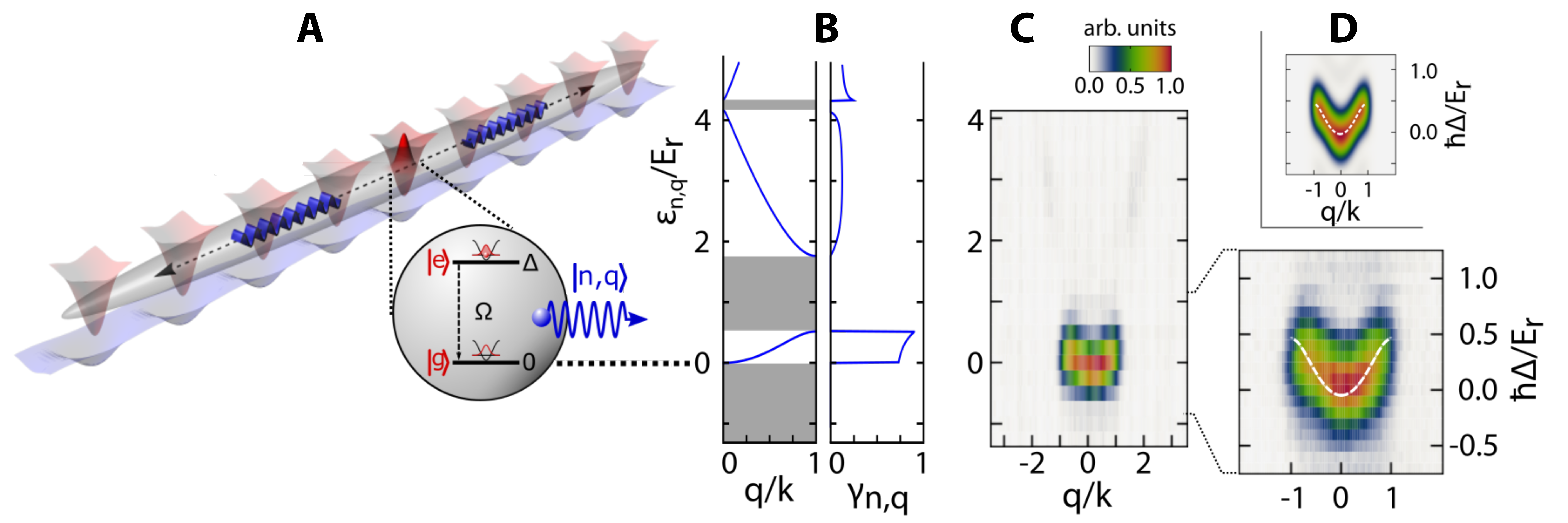}
    \caption{\textbf{A}, Experimental scheme. $^{87}$Rb atoms in two hyperfine ground states $\ket{r}$ (red) and $\ket{b}$ (blue) are confined in state-independent 1D lattice tubes. A state-dependent longitudinal lattice provides strong confinement for $\ket{r}$ ($s_r=20$) and weak confinement for $\ket{b}$ ($s_b\approx2.5$); a coupling between $\ket{r}$ and $\ket{b}$ (strength $\Omega$, detuning $\Delta$) is applied. Each (un)populated $\ket{r}$ well acts as a quantum emitter (states $\ket{\g}$, $\ket{\e}$, excitation energy $\hbar\Delta$) coupled to the band structure of the shallow lattice.
    \textbf{B}, Band structure $\varepsilon_{n,q}$ relevant for the emission of matter waves and relative strength of the vacuum coupling $\gamma_{n,q(\varepsilon)}$  for $s_b=2.5$ and $s_r=20$.
    \textbf{C}, Measured quasimomentum distribution versus emission energy $\hbar\Delta$, as seen with absorption imaging after band-mapping and 14~ms time of flight, and averaged over at least 3 runs. The lattice parameters are as in \textbf{B}; the coupling is applied with strength $\Omega/2\pi = 1.02(3)$~kHz for a duration $\tau=400\mu$s. The zoom in is taken with a smaller step size of 0.1$E_r$, and an average over at least 4 runs for each quasimomentum distribution, and the calculated band-structure is shown (white, dashed).  \textbf{D}, Theoretically computed $|B_q(\tau)|^2$, blurred by a Gaussian of width $\sigma_E=0.1E_r$ in energy and $\sigma_q =0.15k$ in quasimomentum for comparison with the experimental data of \textbf{C} (see Appendix B).     }
\label{FIG:SystemIntro}
\end{figure*}

\section{Implementation} Our experimental scheme, illustrated in Fig.~\ref{FIG:SystemIntro}\textbf{A}, consists of three elements. First, we create a system of isolated lattice tubes confining ultracold, optically-trapped $^{87}$Rb atoms in two relevant hyperfine ground states $\ket{r}=\ket{F=1,m_F=-1}$  and $\ket{b}=\ket{2,0}$ via a deep 2D optical lattice at 1064~nm.  The lattice tubes act as 1D waveguides in which the atoms can freely propagate  (for sufficiently short times $\tau\ll\tau_z=2\pi/\omega_z\approx10$ms) along the tube axis. Second, we create an array of quantum emitters and the structured vacuum by applying an additional state-dependent lattice along the axis of each tube with depths $s_{r}=20$ and $s_b\ll s_r$ (in units of $E_{r}=(\hbar k)^2/2m$, where $m$ is the atomic mass, and $k=2\pi/\lambda$ with $\lambda$ the lattice wavelength); this lattice tightly confines $\ket{r}$ atoms in the harmonic oscillator ground state $\ket{\psi_\e}$ of each well (with size $\ll\lambda$), while it provides a tunable band structure for $\ket{b}$. Finally, we implement a coupling between $\ket{r}$ and $\ket{b}$ states by applying an oscillatory microwave field with strength $\Omega$ and detuning $\Delta$ from the lattice-shifted $\ket{r}\leftrightarrow\ket{b}$ resonance at 6.8~GHz,  thereby inducing transitions between $\ket{r}$ in a well and $\ket{b}$ in a continuum of momentum modes. As a result, each lattice well acts as a quantum emitter of a $\ket{b}$ atom with an excitation energy $\hbar\Delta$ and effective vacuum coupling $\propto\Omega$, where a populated lattice well takes on the role of the emitter's excited state $\ket{\e}$ and an unpopulated well plays the role of its ground state $\ket{\g}$. Our experiments start with a sparsely and incoherently populated lattice (filling fraction $\lesssim 0.5$), so that a majority of the quantum emitters are in the ground state (for details, see Appendix A).

The dynamics of a quantum emitter coupled to the band structure $\varepsilon_{n,q}$ of the shallow lattice (with band index $n$ and quasimomentum $q\in[-k,k]$) is then described by the Weisskopf-Wigner type Hamiltonian $\hat{H} = \sum_{n, q} \hbar g_{n,q} e^{i\Delta_{n,q} t} \ket{\g}\bra{\e}\hat{b}^\dagger_{n,q} + \text{H.c.}$, where $\Delta_{n,q} = \Delta-\varepsilon_{n,q}/\hbar$ is the effective detuning of the emitter from the Bloch state  $\ket{n,q}=\hat{b}^\dagger_{n,q}\ket{0}$, and the effective vacuum coupling $g_{n,q} = \gamma_{n,q} \Omega/2$ contains the Franck-Condon overlap $\gamma_{n,q} = \langle n,q|\psi_\e\rangle$.

While the free-space case $s_b=0$ \cite{Krinner2018} corresponds to optical emission in the vicinity of a photonic band edge, a band structure featuring multiple such edges \cite{Oberthaler2006RMP} as in a photonic crystal or a coupled-cavity array can readily be implemented by tuning $s_b$ via $\lambda$.
For our measurements, we generally choose $s_b = 2.5$ (at $\lambda = 790.4$~nm) for which the width of the ground band is $ \varepsilon_{1,k}-\varepsilon_{1,0} = 0.5~E_{r}\approx h\times 1.8$~kHz. The band structure and corresponding couplings for these parameters are illustrated in Fig.~\ref{FIG:SystemIntro}\textbf{B}.

\section{Bloch-wave emission spectrum} To access effects of the band structure, we first measure the momentum distribution of the emitted $\ket{b}$ atoms as a function of the excitation energy $\hbar\Delta$. For this purpose, we apply a rectangular microwave pulse of duration $\tau=400\mu$s and Rabi frequency $\Omega=2\pi\times1$ kHz, which is then followed by a $500\mu$s-long ramp down of all three lattices for the purpose of band mapping. The emitted $\ket{b}$ atoms are then detected after time-of-flight using state-selective absorption imaging. The measured quasimomentum distribution, shown in Fig.~\ref{FIG:SystemIntro}\textbf{C}, is very different from the parabolic shape seen for free-space emission \cite{Krinner2018}, and clearly reveals the presence of a gapped spectrum. In addition, the  emission into the ground band is seen to be much stronger than that into the first and higher bands. This suppression  can be explained by the structure of the vacuum coupling $g_{n,q}$, which for even-$n$ bands and $s_b>0$ is reduced due to the approximate odd parity of the relevant Bloch states (as opposed to the case $s_b<0$; cf. Appendix A); a further suppression for higher $n$ is due to the finite momentum width of $\ket{\psi_\e}$ and the decrease in the density of states. As a result, our system is closely modeled by a single-band picture in which all the dynamics is induced by coupling to the ground band. For our parameters, the band  is approximately sinusoidal, $\varepsilon(q) = - \hbar\bar{\omega}\cos(q \pi/k)+\hbar\bar{\omega}$ (denoting $\varepsilon(q)\equiv\varepsilon_{1,q}$, and $\hbar\bar{\omega}\equiv (\varepsilon_{1,k}-\varepsilon_{1,0})/2 = 0.25 E_r$) and the vacuum coupling $g = \langle g_{1,q}\rangle_q \approx 0.39\Omega$ is approximately constant over the band.

\section{Band decay} 
\begin{figure}[th!]
\centering
\includegraphics[width=1\columnwidth]{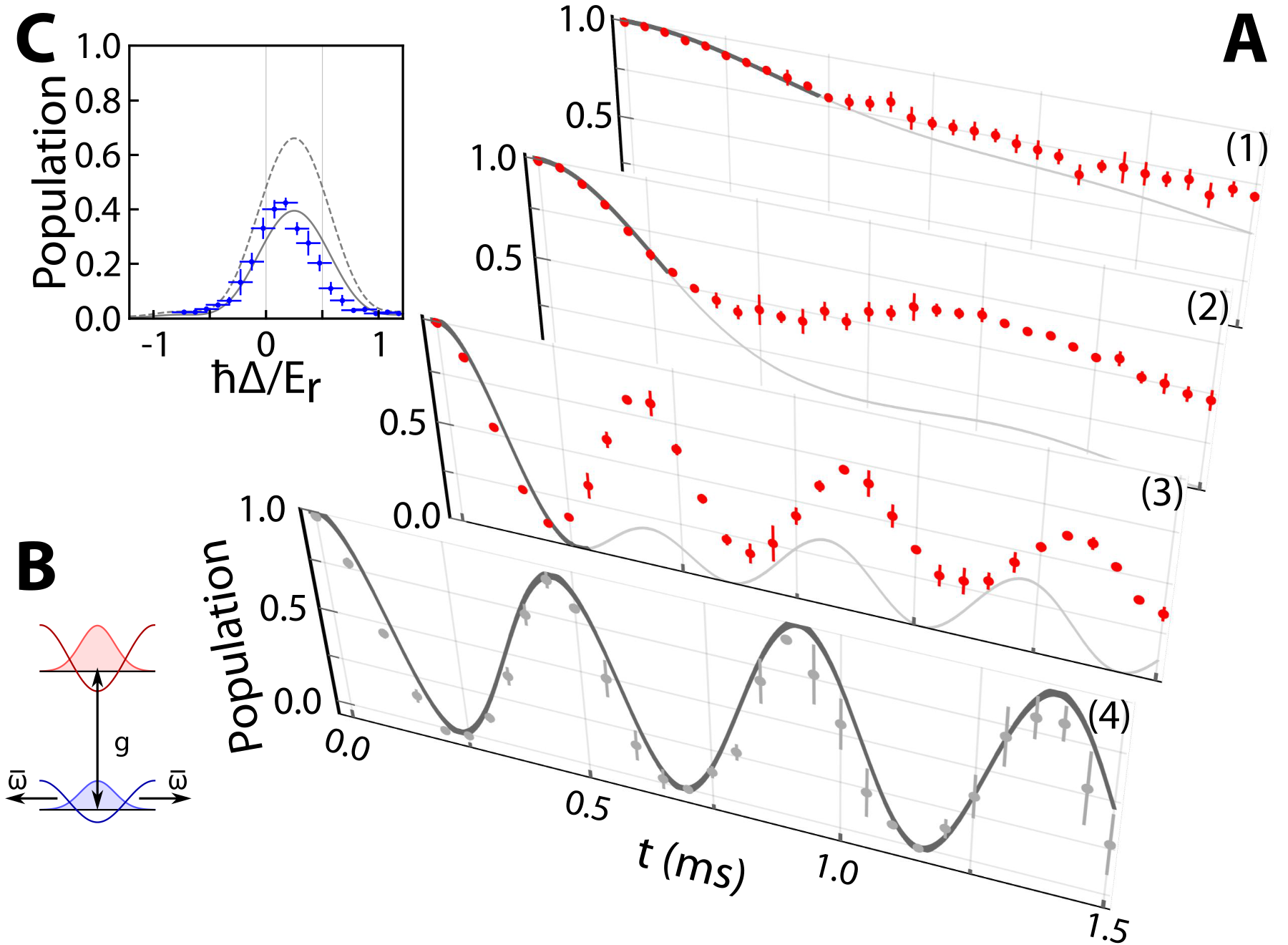}
    \caption{\textbf{A}, Decay dynamics at the band center $\Delta = \bar{\omega}$ with (1) weak coupling with $\Omega = 2\pi\times0.4$~kHz $(g/\bar{\omega} = 0.18)$, (2,3) intermediate coupling with $\Omega = 2\pi\times(1.0,2.3)\mbox{~kHz}$  ($g/\bar{\omega} = 0.43,1.0$), and (4) strong coupling for a reduced bandwidth ($0.1E_r$) with $\Omega=2\pi\times2.2$~kHz ($g/\bar{\omega}=4.9$). The dots are data taken for different hold times, averaged over at least 3 runs each, with error bars representing the standard error of the mean. The gray lines represent the predictions of the isolated-emitter model, with its estimated range of applicability indicated in bold. \textbf{B},  Schematic illustrating the competition between coupling $g$ and tunneling $\sim\bar{\omega}$ in the shallow lattice (see text). \textbf{C}, Emission spectrum for $g/\bar{\omega}=0.43$ and $\tau=400~\mu$s, obtained from the distributions in Fig. 1 \textbf{C} by summing over quasimomenta. The dashed curve is the prediction of the single-emitter model, and the solid curve is the same prediction reduced by 40\%.
    }
\label{FIG:TimeEvolCompare}
\end{figure}

The  dynamics inside the ground band depends on the relative strength of the vacuum coupling and the bandwidth. With the excitation energy tuned to the center of the band, we monitor the time evolution of the excited-state amplitude for different ratios $g/\bar{\omega}$; the results are shown in ~Fig.~\ref{FIG:TimeEvolCompare}\textbf{A}. Three regimes can be distinguished: irreversible decay for $g/\bar{\omega}\ll1$ (weak coupling, ~Fig.~\ref{FIG:TimeEvolCompare}\textbf{A}(1)), damped oscillatory decay for $g/\bar{\omega}\sim1$ (intermediate coupling, ~Fig.~\ref{FIG:TimeEvolCompare}\textbf{A}(2,3)), and undamped oscillations for $g/\bar{\omega}\gg1$ (strong coupling, ~Fig.~\ref{FIG:TimeEvolCompare}\textbf{A}(4)).

For an isolated emitter, a Wannier picture provides a qualitative description of the coupling dependence in terms of the quantum Zeno effect \cite{Peres1980,Itano1990} (cf. Fig.~\ref{FIG:TimeEvolCompare}\textbf{B}): here, the atom coherently cycles with Rabi frequency $2g$ between the strongly confining emitter well and a corresponding $\ket{b}$ well of the shallow lattice, where it is subject to tunnel escape at a rate $\sim\bar{\omega}$ that damps the coherent local evolution, with exponential decay for weak coupling $g/\bar{\omega}\ll1$. Switching back to the band picture, in this case the band edges are both far away in energy such that the situation becomes analogous to spontaneous free space decay \cite{Weisskopf1930}. On the other hand, for strong coupling the bandwidth becomes negligible, and the cavity-QED limit with an effectively single-mode vacuum is recovered (similarly, and independent of the coupling strength, the initial dynamics is Rabi-like for times $t\lesssim\bar{\omega}^{-1}=0.17$~ms for which the associated Heisenberg uncertainty in energy exceeds the bandwidth \footnote{Similar behavior has been observed in quantum tunneling experiments \cite{Wilkinson1997}.}).

While an analytical treatment of the dynamics of a multiple-emitter system is beyond the scope of this work, an isolated emitter is described by the interaction-picture Schr\"odinger equation for $\hat{H}$ with the ansatz  $\ket{\psi(t)} = A(t)\ket{\e; 0}+\sum_q B_q(t)\ket{\g; 1,q}$, where $A(t)$ is the excitation amplitude (with $A(0)=1$) and $B_q(t)$ the spectral amplitude of the matter-wave radiation field. The analogue scenario for photonic crystals has previously been analyzed \cite{Lombardo2014,Calajo2016,Gtudela2017B}, yielding $A(t) = \frac{i}{2\pi}\int_{-\infty}^\infty \operatorname{d}\hspace{-1pt}\omega G(\omega+i0^+) e^{i(\Delta-\omega) t}$, where $G(\omega) = 1/[\omega-\Delta-\Sigma(\omega)/\hbar]$ is the  Green's function containing the self-energy  $\Sigma(\omega) =-i\hbar g^2/\sqrt{\omega(2\bar{\omega}-\omega)}$, which captures the backaction of the band on the emitter. From this result for $A(t)$, it is then straightforward in our case to obtain the spectral amplitudes $B_q(t) = -i g\int_0^t\operatorname{d}\hspace{-1pt}\tau e^{i(\varepsilon(q)/\hbar-\Delta)\tau}A(\tau)$ of the emitted matter waves through numerical integration  (cf. Appendix B). In the following, we will compare the predictions of this simplified model to observed decay and emission behavior.

The model's predictions for $A(t)$  are shown in  Fig.~\ref{FIG:TimeEvolCompare}\textbf{A} alongside the data for decay in the band center. There is indeed good agreement before significant population ($>10\%$) leaves the Wigner-Seitz cell of the emitter considered (cf. Appendix B), validating the applicability of the isolated-emitter model in this regime. More generally, the agreement between the observed dynamics and the model degrades as time progresses (with the exception of $g/\bar{\omega}\gg1$, where the emitters are effectively decoupled).  The observed deviations such as an offset and enhanced oscillations are qualitatively similar to those already seen in free-space emission \cite{Krinner2018} which arise from coupling to neighboring, initially empty emitters. Other possible effects, arising from the weak longitudinal confinement or collisional interactions, are expected to be less significant on the observed time scales.

To test the model's predictions for the dependence on excitation energy, we compute $|B_q(\tau=400\mu\mbox{s})|^2$ as a function of $\hbar\Delta$, with $q$ limited to the first Brillouin zone, and compare it to the measured ground-band momentum distribution of Fig.~\ref{FIG:SystemIntro}\textbf{C}. With a small amount of blurring due to magnetic-field jitter and the finite size of the sample, the results of the calculation, shown in Fig.~\ref{FIG:SystemIntro}\textbf{D}, closely resemble the  momentum features of the data. Moreover, the model cleanly reproduces the integrated spectrum, cf.~Fig.~\ref{FIG:TimeEvolCompare}\textbf{C}, up to an overall scaling factor of order unity consistent with the time evolution.

\section{From dressed to bound states} The Rabi oscillations for  $g/\bar{\omega}\gg1$ involve emitted $\ket{b}$ atoms in the Wannier functions of the shallow lattice, featuring equally strong contributions from all Bloch waves of the band. The Wannier functions are fixed by the lattice potential and together with $\ket{\psi_\e}$ form dressed states as in cavity QED.
This picture breaks down for $g/\bar{\omega}\sim1$ when the band is spectrally resolved such that certain quasimomenta contribute more strongly than others, with the consequence that the spatial shape of the emitted radiation becomes dependent on both $g$ and $\Delta$. In generalization of the free-space case \cite{Stewart2017} the two dressed states are replaced by two bound states in which the $\ket{b}$ atoms form an evanescent wave around the emitter (taking the role of the Wannier function), which is given by $\psi_{B}^\pm(z) = \int_{-k}^k \operatorname{d}\hspace{-1pt}q~ \phi_B^\pm(q) \ip{z}{1,q}$, with quasimomentum probability amplitudes $\phi_B^\pm(q) = (\hbar g/2k)/[\hbar\omega_B^\pm-\varepsilon(q)]$ (for details, see Appendix B). The bound-state energies $\hbar\omega_B^\pm$, which are obtained as real-valued poles in $G(\omega)$ and vary with  $\Delta$, $g$, and $\bar{\omega}$, are outside of the band, and converge to the band edges from above and below for decreasing $g/\bar{\omega}$. In addition to the real poles, which lead to Rabi-like oscillations with reduced amplitude, $G(\omega)$ supports other singularities \cite{Gtudela2017B} which are responsible for Markovian and non-Markovian decay of the emitter population  as the ratio $g/\bar{\omega}$ is varied.

\begin{figure}[t!]
\centering
    \includegraphics[width=1\columnwidth]{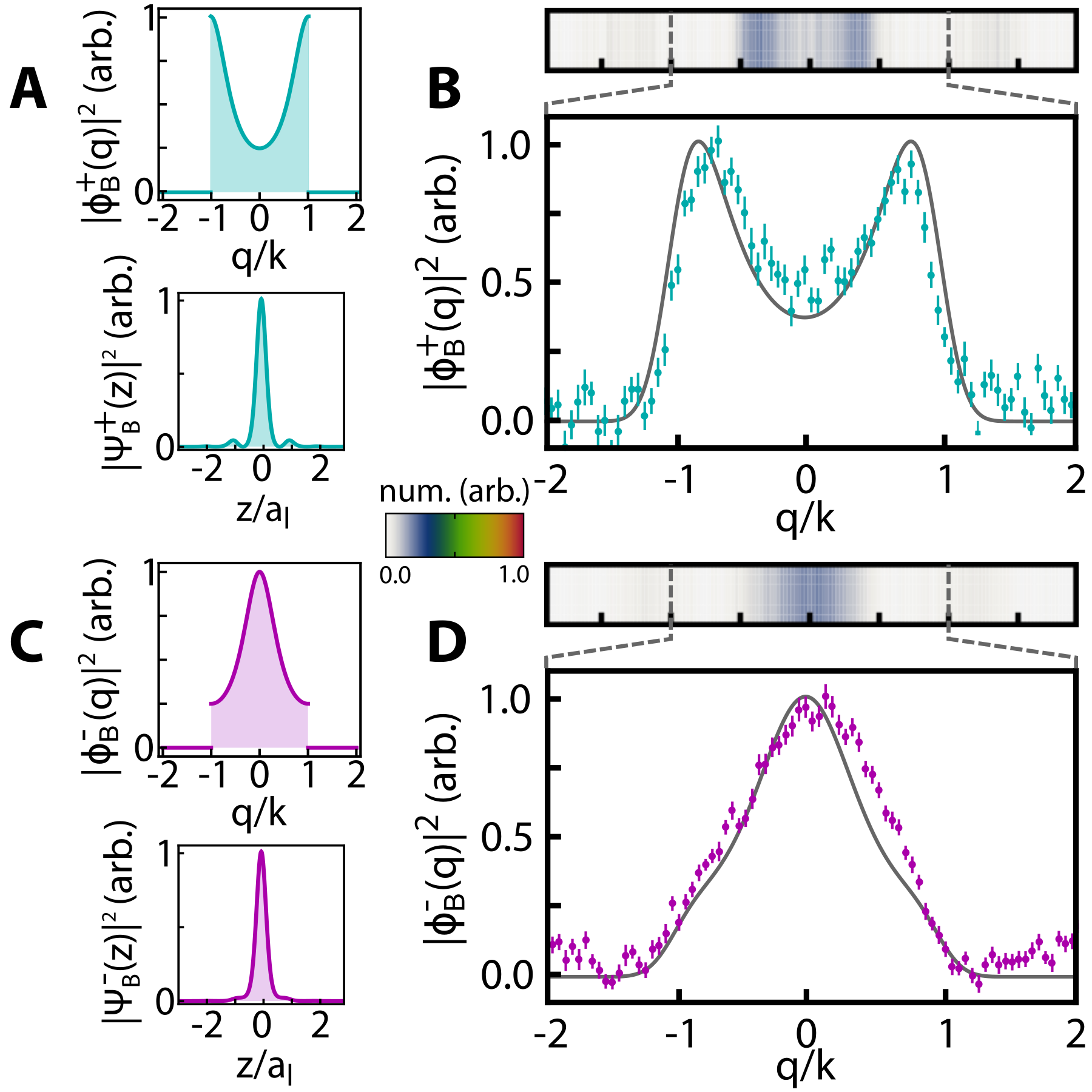}
    \caption{Structure of the bound states at $\hbar\Delta^+ = 1.0(1) E_r$ and $\hbar\Delta^- = -0.5(1)E_r$ above (\textbf{A}, \textbf{B}) the upper, and below  (\textbf{C}, \textbf{D}) the lower band edge.
    \textbf{A}, Calculated quasimomentum distribution $|\phi_B^\pm (q)|^2$ of the bound state above the band at $\hbar\omega_B^+ = 4\hbar\bar{\omega} \approx h\times3.9(3)$ kHz and corresponding computed position space distribution $|\psi^\pm (z)|^2$, where $a_l = \lambda/2$ is the lattice spacing.
    \textbf{B}, Observed quasimomentum distribution from time-of-flight following an adiabatic ramp on (2 ms long) of the coupling up to $g/\bar{\omega} = 0.43$. Each data-point is the average of more than 30 individual repetitions. The density plot shows the average time-of-flight picture. The gray curve is the quasimomentum distribution from \textbf{A} convolved with a Gaussian blur ($\sigma_q = 0.15k$) to accommodate finite size effects and imaging resolution. \textbf{C}, Quasimomentum and position distributions as in \textbf{A} for the bound state below the band at $\hbar\omega_B^- = -2\hbar\bar{\omega} \approx h\times-1.9(3)$ kHz; the latter exhibits a small plateau for our parameters.  \textbf{D}, observed quasimomentum distribution, taken as in \textbf{B}. The gray curve is blurred with the same Gaussian as in panel \textbf{B}.
    }
\label{FIG:BoundStateShapes}
\end{figure}

In order to compare with the model, we directly access two representative bound states located on opposite sides of the band, $\hbar\Delta^\pm=(1\pm3)\hbar\bar{\omega}$, and weak coupling $|g/\Delta^\pm|\ll1$ (such that $\omega^\pm_B\approx\Delta^\pm$). To avoid non-adiabatic emission effects \cite{Krinner2018}, we prepare these states by slowly ramping on the coupling $g$ using a sinusoidal ramp. The ramp duration of 2 ms is long with respect to the
bound state frequencies $\omega^\pm_B$, and no dynamics are observed for a variable hold time between 0 and 0.5 ms following the ramp, confirming that the resulting state is stationary. The resulting quasimomentum distributions are observed in time-of-flight after a band-map of all optical potentials as before. The observed distributions, cf. Fig.~\ref{FIG:BoundStateShapes}\textbf{B} and \ref{FIG:BoundStateShapes}\textbf{D}, match qualitatively the predictions for $|\phi_B^\pm(q)|^2$ within the range $-2k$ to $2k$, with quantitative agreement if we allow for a blurring of $0.15 k$ due to finite size effects (system size $\sim10\mu$m) and imaging resolution. We note that the two states are copies of each other displaced by half of the Brillouin zone, $\phi_B^\pm(q+k)=\phi_B^\mp(q)$.

The spatial profile of the lower bound state ($\psi_B^-$) is similar to that found below a continuum with a single edge \cite{Krinner2018} and the form usually considered in the literature (albeit with quasimomentum cutoffs at $\pm k$ that lead to a slight modification of its exponential decay \cite{Bykov1975}). In contrast, the strong contributions from $q=\pm k$ in the upper bound state ($\psi_B^+$) lead to strong deviations from exponential localization, with strong modulations at the lattice period featuring the nodes of a standing wave.

\section{Conclusions.} In this work, we have shown that decay in a band presents new features not present in free-space matter-wave emission. In particular, fractional decay changes its character to longer-time oscillations when a second bound state is present. In the context of open quantum systems, these oscillations represent a partial retrieval of information lost to the environment at well-defined times, which is not realizable with only one bound state. These bound states also provide insight into the corresponding states in photonic band-gap materials, where our results might be relevant for the engineering of  bound-state-mediated  long-range interactions \cite{Hood2016} as they highlight the importance of the positioning of the quantum emitter with respect to a photonic crystal, where long-range couplings between emitters may be susceptible to small displacements on the scale of the lattice period. The accessibility of higher bands and tunable geometries will provide flexibility for studies of effective spin-models, analogs of chiral emission, and collective emission phenomena \cite{Vega2008,GTudela2018B,GTudela2017}.

\section{Acknowledgements.}
We thank Y. Kim and M. G. Cohen for discussions and a critical reading of the manuscript. This work was supported by NSF (Grants No. PHY-1607633 and No. PHY-1912546). M.S. was supported by a GAANN fellowship from the US Department of Education, and A.L. received support, partially from a Spain-US Fulbright grant co-sponsored by the  Ram\'on Areces Foundation and partially from SUNY Center for Quantum Information Science on Long Island.

\appendix
\section{Experimental procedures}

\paragraph*{\it Sample preparation:}
The experiment begins by creating an optically-trapped Bose-Einstein condensate \cite{Pertot2009supp}. In order to minimize gravitational sag, the horizontal, state-independent lattices are first adiabatically ramped up in 80~ms followed by the vertical state-dependent lattice (90 ms) to final depths of $40E_{r,1064\text{nm}}, 40E_{r,1064\text{nm}}$ and $20E_{r,790.41\text{nm}}$ so that the atomic cloud sits at approximately the trap minimum potential, with a residual confinement along the $z$-direction of $\omega_z\approx2\pi\times100$ Hz. Here, $E_{r,\lambda}$ is the recoil energy of the lattice. This procedure creates an atomic sample deep within the Mott regime.
With the atoms loaded into the lattice, a variable fraction $f$ is then transferred, at a bias field of 5 G, to an intermediate $\ket{F=2,m_F=1}$ state using a two-photon microwave and radio-frequency pulse of about 2 ms duration.
The transferred atoms are removed using resonant light on the   $D_2$ cycling transition ($F=2\to F^\prime=3$). After the pulse sequence (in which $f$ is adjusted between 0.6 and 0.85 to compensate for differing initial atom number), the remaining sample has about $2.7(3)\times10^4 \ket{r}$ atoms with an average site occupation of $\ev{n_i}\lesssim0.5$ in the tubes.

\paragraph*{\it State-dependent lattice and atom detection:}
Our experimental techniques follow that of our previous work \cite{Krinner2018}. In brief, we generate the state dependent potential using $\sigma^-$ light tuned to $\lambda = 790.4$ nm, between the  $D_1$ and $D_2$ transitions of $^{87}$Rb. We detect the atoms after a 500 $\mu$s-long linear ramp-off of all optical lattice potentials to perform a band-map operation followed by 14~ms of time-of-flight (ToF) expansion. During ToF, we apply Stern-Gerlach separation using a magnetic field gradient in order to spatially separate hyperfine states of different magnetic moment. We then perform state-selective absorption imaging in order to resolve all hyperfine states in each ground state manifold individually (used for magnetometry \cite{Krinner2018Bsupp}). Images are analyzed for data extraction after using a principal component analysis routine to remove residual fringes in the images.

\paragraph*{\it Determining the resonance condition:}
The resonance condition $\Delta=0$ is defined with respect to the transition between the band minimum $\varepsilon_{n=1,q=0}$ and the harmonic-oscillator ground state in the matter-wave emitter potential (with a residual bandwidth of $0.01~E_r$). We use lattice transfer spectroscopy \cite{Reeves2015supp} to determine the resonance condition. An optically trapped BEC of $\ket{r}$ atoms is first transferred into the $\ket{b}$ state, after which the state-dependent lattice potential is ramped on slowly. Microwave pulses of duration $\tau=400\mu$s are then applied at a fixed strength $\Omega=2\pi\times 1.0$~kHz and variable frequency to transfer maximally 30\% of population into the $\ket{r}$ state. The $\Delta = 0$ frequency for use in the experiment is obtained from a fit of a Rabi spectrum to the data. Systematic residual mean-field shifts are estimated to be between $150$ and $270$~Hz for all initial atom numbers used, based on a direct simulation of the 1D time-dependent Gross-Pitaevskii equation, and have been included in the spectrum of Fig. 2\textbf{C}. The resonance condition (which depends on both optical and magnetic fields) is stabilized using a post-selection magnetometry technique, yielding an uncertainty of $\sigma_E\approx h\times$~350~Hz $\approx 0.1 E_r$ \cite{Krinner2018Bsupp, Krinner2018}.

\paragraph*{\it Higher-band contributions:}
 The observed quasimomentum distributions show a small ($\lesssim20$\%) population of atoms at higher quasimomenta ($q\approx 2.5 k$).
 This can be attributed to a small contamination by the first excited harmonic-oscillator level for $\ket{r}$ at the beginning of the measurement which is coupled to the first excited band for the $\ket{b}$ atoms due to a non-vanishing Franck-Condon overlap. These atoms are in a different region of quasimomentum space from the evanescent waves in the experimental data.

 \begin{figure}[H]
\centering
    \includegraphics[width=0.9\columnwidth]{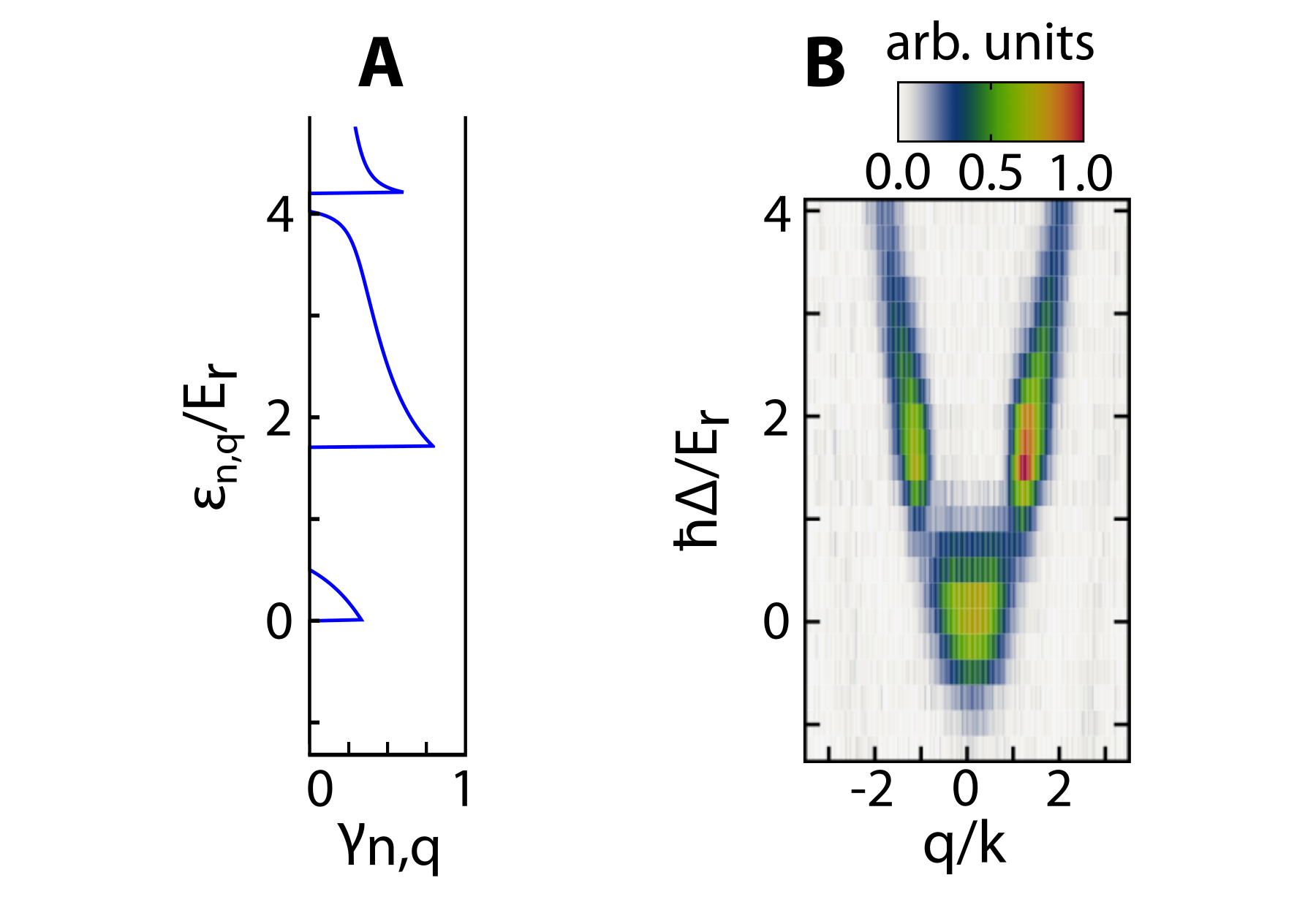}
    \caption{\textbf{A}, Franck-Condon factor $\gamma_{n,q}$ and observed emission profile \textbf{B}, for $s_b=-2.6$ at $\lambda = 789.8$ nm, with all other parameters as in Fig.~1\textbf{C}. The strongest emission signal occurs in the first excited band; the relatively strong percentage of atoms within the first band gap results from the strong coupling to the first excited band, giving rise to non-Markovian emission in the band gap.}
\label{FIG:DisplacedCase}
\end{figure}

\paragraph*{\it Positioning of the emitters:}
In the experiments in the main text, the commensurability of the emitter array with the shallow lattice creating the band structure guarantees that the coupling is uniform across the sample. Furthermore, it is possible to realize the case  $s_b<0$  by shifting the lattice wavelength in the opposite direction from the tune-out point. Physically, this corresponds to shifting the emitter lattice with respect to the shallow lattice by half of a lattice spacing, aligning the emitters with the unit cell boundaries. This results in Franck-Condon factors that are appreciable for excited bands $n\geq2$, cf. Fig.~\ref{FIG:DisplacedCase}\textbf{A}, and in observed emission profiles with appreciable contributions from both the ground and first exited bands, cf. Fig.~\ref{FIG:DisplacedCase}\textbf{B}.

\section{Theoretical considerations}

\paragraph*{\it Time evolution:}
The time evolution of the excited emitter population is determined by solving $A(t) = (i/2\pi)\int_{-\infty}^\infty \operatorname{d}\hspace{-1pt}\omega G(\omega+i0^+) e^{i(\Delta-\omega) t}$ using the techniques of complex analysis. As demonstrated in \cite{Gtudela2017B} (see also \cite{Lombardo2014,Calajo2016}), there are three kinds of singularities in $G(\omega)$ which contribute: stable poles corresponding to bound states outside the band, an unstable pole inside the band, and an incoherent loss due to branch cuts at the band edges. The equations of motion can then be solved numerically for any desired parameters by solving for the poles with their residues and numerically integrating along the branch cuts.

\paragraph*{\bf Numerical computation of bound states.}
By using the method of Laplace transforms and assuming a stable bound-state pole, we can find the composition of the bound state. Specifically, using the same steps as in Ref. \cite{Stewart2017} (cf. Eqs.(37-38) and (40-46) therein), one finds
\begin{equation}
    \label{eq:bound_state_def}
    \ket{\Psi^\pm_B} = N\left(\ket{\e;0}+\frac{g}{2k}\int_{-k}^{k} \operatorname{d}\hspace{-1pt}q \frac{\ket{\g;1,q}}{\omega^\pm_B-\varepsilon(q)/\hbar}\right),
\end{equation}
with $N$ a normalization constant. In the limit of strong coupling $g/\bar{\omega}\gg1$, the second term reduces effectively to a ground state emitter and a Wannier function in the shallow lattice $\ket{\g;\text{w}_b}$ commensurate with the position of the emitter, with $\omega_B^\pm = \pm g$ dominating the integral. Thus, we see that $\ket{\Psi_B^\pm} = N(\ket{\e;0}\pm\ket{\g;\text{w}_b})$ maps to the dressed states of the Jaynes-Cummings model in this limit.

In order to learn about the spatial shape of the emitted radiation component, we compute numerically (by exact diagonalization) the Bloch waves, $\psi_q(z)$, corresponding to the band structure of interest with small quasimomentum spacing (100 steps across the Brillouin zone) and add the results according to the defining equation $\psi^\pm_{B}(z) = g\int_{-k}^{k}\frac{\operatorname{d}\hspace{-1pt}q}{2k} \psi_q(z)/[\omega_B^\pm-\varepsilon(q)/\hbar]$.

\begin{figure}[h!]
\centering
    \includegraphics[width=0.9\columnwidth]{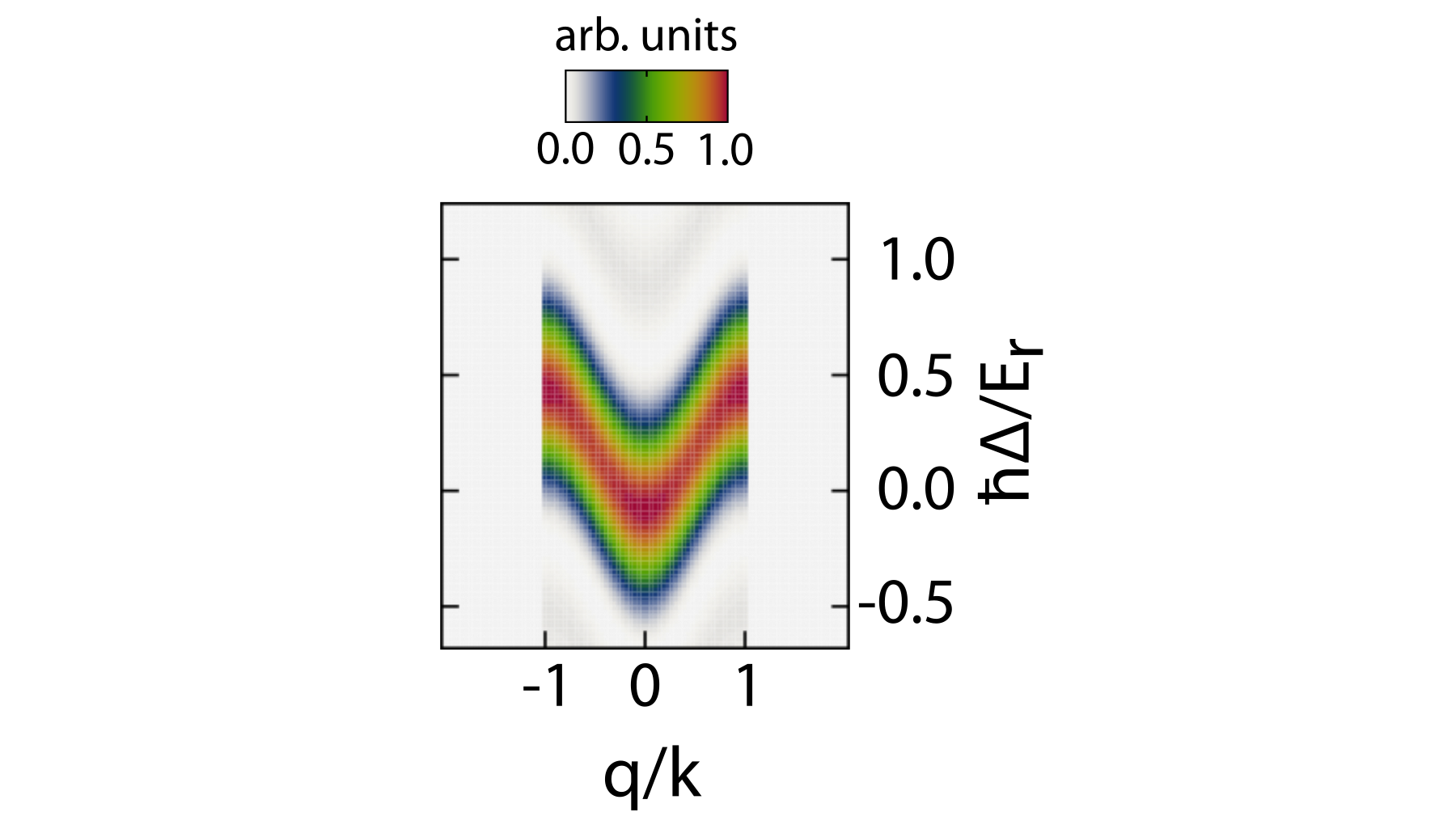}
    \caption{Computed $|B_q(t)|^2$ for $-k\leq q \leq k$. The single band model predicts identically zero contribution outside of the first Brillouin zone.
    }
\label{FIG:UnblurTheory}
\end{figure}

\paragraph*{\it Theoretical computation of $B_q(t)$:}
In order to calculate $|B_q(t)|^2$ for $t=400\mu$s, we integrate
$B_q(t) = -i g\int_0^t\operatorname{d}\hspace{-1pt}\tau e^{i(\varepsilon(q)/\hbar-\Delta)\tau}A(\tau)$,
or after swapping the order of integrations,
\begin{equation}
    \label{eq:bq_2}
    B_q(t) = \frac{i g}{2\pi}\int_{-\infty}^\infty \operatorname{d}\hspace{-1pt}\omega G(\omega + i0^+) \frac{ e^{i(\varepsilon(q)/\hbar-\omega)t}-1}{ \omega+i0^+-\varepsilon(q)/\hbar},
\end{equation}
for the first Brillouin zone, and set $|B_q(t)|^2$ equal to zero outside this zone, cf. Fig.~\ref{FIG:UnblurTheory}. This reflects the band-mapping procedure arranging the quasimomenta in an extended-zone scheme and our model having only one band. In order to make a comparison with the experimental data of Fig.~1\textbf{C}, we apply a Gaussian blur of $\sigma_E = 0.1E_r$ in the energy axis and $\sigma_q = 0.15 k$ in the momentum axis to account for magnetic field and finite size uncertainty.\\

Just as in the case of $A(t)$, one can apply the residue theorem in~\eqref{eq:bq_2} to split the emission into a part due to stable poles (bound states) and a decaying part (unbounded emission), $B_q(t)=B_q^{B}(t)+B_q^{de}(t)$. The decaying part present at the $n^{\mathrm{th}}$ Wigner-Seitz cell $B_n^{de}(t)=\int_{-k}^k \frac{\operatorname{d}\hspace{-1pt}q}{2k}B^{de}_q(t)\cos(n\pi q)$ allows estimating the influence of neighboring emitters in the dynamics. More specifically, they start to play a role when the unbounded emission outside the original Wigner-Seitz cell is about $10\%$ ($\sum_{n\neq0}\left\lvert B^{de}_n(t) \right\rvert^2\gtrsim0.1 $).

\normalem

\end{document}